\documentstyle[12pt]{article}
\def\gsim{\stackrel{>}{\sim}}
\def\lsim{\stackrel{<}{\sim}}

\def\beq{\begin{equation}}
\def\eeq{\end{equation}}
\def\ul{\underline}
\def\eg{$e. g. \,$}
\def\ie{$i. e. \,$}
\def\ave{<\!\sigma_{{\rm ann}}\!>}
\def\nuave{<\!\! n_{\nu_j}\!\!>}
\def\zb{$Z$--burst}
\def\epm{$e^{\pm}$}
\def\enn{<\!N\!>}
\def\h100{h_{100}}
\begin{document}
%
%
\begin{centering}
{\large{\bf Cosmic--Ray Neutrino Annihilation on Relic Neutrinos Revisited: 
A Mechanism for Generating  Air Showers above the  
Greisen--Zatsepin--Kuzmin Cutoff}}\\
\vspace{1cm}
%
Thomas J. Weiler\\ 
%
{\it Department of Physics \& Astronomy, Vanderbilt University, 
Nashville, TN 37235}\\
and\\
{\it Department of Physics, University of Wisconsin,
Madison, WI 53706}\\
\vspace{0.35cm}
email: weilertj@ctral1.vanderbilt.edu\\
\end{centering}
\vspace{0.5cm}
\begin{abstract}
\noindent
If neutrinos are a significant contributor to the matter density
of the universe, then they should have $\sim$ eV mass and cluster
in galactic (super) cluster halos, 
and possibly in galactic halos as well. 
It was noted in the early 1980's that cosmic ray neutrinos with energy 
within $\delta E/E_R=\Gamma_Z/M_Z \sim~3\%$ of the peak energy
$E_R=4\;({\rm eV}/m_{\nu})\times 10^{21}$ eV will annihilate
on the nonrelativistic relic antineutrinos (and vice versa) to 
produce the $Z$--boson with an enhanced, resonant cross section of
${\cal O}(G_F)\sim 10^{-32}{\rm cm}^2$.
The result of the resonant neutrino annihilation is
a hadronic $Z$--burst 70\% of the time, 
which contains, on average, thirty  photons  and 
2.7 nucleons with energies near or above the GZK cutoff energy of 
$5\times 10^{19}$ eV.  
These photons and nucleons produced within our Supergalactic halo 
may easily propagate to earth and initiate
super--GZK air showers.
Here we show that the probability for each
neutrino flavor at its resonant energy to annihilate 
within the halo of our Supergalactic cluster is 
likely within an order of magnitude of 1\%, 
with the exact value depending on unknown aspects
of neutrino mixing and relic neutrino clustering.
The absolute lower bound in a hot Big Bang universe 
for the probability to annihilate within a 50 Mpc radius 
(roughly a nucleon propagation distance) of earth is 0.036\%.
From fragmentation data for $Z$--decay, 
we estimate that the nucleons are more energetic than the photons
by a factor $\sim 10$.
Several tests of the hypothesis are indicated.
\end{abstract}
\vfill\eject
\section{The Cosmic Ray Puzzle Above $10^{20}$~eV}
The recent discoveries by the AGASA\cite{akeno}, Fly's Eye\cite{eye}, 
Haverah Park\cite{hp}, and Yakutsk\cite{yak} collaborations
of air shower events with energies above the Greisen--Zatsepin--Kuzmin (GZK) 
cutoff  of $\sim 5\times 10^{19}$ eV presents an outstanding puzzle in 
ultrahigh--energy cosmic--ray physics. 
It was anticipated that the highest--energy cosmic primaries would be protons 
from outside the galaxy, perhaps produced in active galactic nuclei (AGNs).
It was also anticipated that the highest energies for protons arriving at earth
would be $\sim 5\times 10^{19}$ eV\cite{gzk}.
The origin of this GZK cutoff is degradation of the proton energy by the 
resonant scattering process 
$p+\gamma_{2.7K}\rightarrow\Delta^*\rightarrow  N+\pi$
when the proton is above the resonant threshold for $\Delta^*$ production;
$\gamma_{2.7K}$ denotes a photon in the $2.7K$ cosmic background radiation.
For every mean free path $\sim 6$ Mpc of travel,
the proton loses 20\% of its energy on average\cite{hs}.
A proton produced at its cosmic source with an initial energy $E_p$ 
will on average arrive at earth with only a fraction 
$\sim (0.8)^{D/6\,{\rm Mpc}}$ of its original energy.  
Since AGNs are hundreds of megaparsecs away, the energy requirement 
at an AGN source for a proton which arrives at earth with a super--GZK energy 
is unrealistically high\cite{AGNmax}. 
Of course, proton energy is not lost significantly if the highest energy 
protons come from a rather nearby source,
$\lsim 50$ to 100 Mpc\cite{50Mpc}.
However, no AGN sources are known to exist within 100 Mpc of earth.
Hence, the observation of air shower events above $5\times 10^{19}$ eV
challenges standard theory.\footnote 
{The suggestion has been made
that hot spots of radio galaxies in the Supergalactic plane 
at distances of tens of megaparsecs may be the sources of 
the super--GZK primaries\cite{stanev}.
}
In principle, there exists a check of the cutoff dynamics.
The expected energy losses for protons
above the cutoff should lead to an event pile--up at energies 
just below the cutoff\cite{hillschramm}.
At present, the data is too sparse to rule in or out an
event pile--up at $\sim 5\times 10^{19}$ eV.

A primary nucleus mitigates the cutoff problem (energy per nucleon
is reduced by 1/A), but has additional problems:
above $\sim 10^{19}$ eV nuclei
should be photo--dissociated by the 2.7K background\cite{stecker}, 
and possibly disintegrated by the 
particle density ambient at the astrophysical source.  
Gamma--rays and neutrinos are other possible primary candidates 
for the highest energy events.
The gamma--ray hypothesis appears inconsistent
with the time--development of the Fly's Eye event,
but is not ruled out for this event\cite{hs}.
However, the mean free path for a $\sim 10^{20}$ eV photon to annihilate
on the radio background to $e^+ e^-$ is believed to be only
$10$ to $40$ Mpc\cite{hs}, and the density profile of the 
Yakutsk event\cite{yak}
showed a large number of muons which may argue against gamma--ray initiation.
Concerning the neutrino hypothesis, 
the Fly's Eye event occured high in the atmosphere,
whereas the expected event rate for early development of a neutrino--induced 
air shower is down from that of an 
electromagnetic or hadronic interaction by six orders of magnitude\cite{hs}.
On the other hand, there is good evidence that 
some of the highest--energy primaries have common arrival directions,
with arrival times displaced by $\sim 10^8$~s.
Such event--pairing argues for a common source of some duration
emitting stable neutral primaries\cite{AGASApairs}.
Neutrino primaries satisfy this criterion.
Charged--particle primaries should display 
bending in cosmic magnetic fields, leading to an energy--dependent 
divergence in arrival directions and arrival times.

Some exotic particles\cite{glueballino,exparts} 
and exotic dynamics\cite{exdynams} 
have been introduced in an attempt to 
circumvent the problems of conventional primaries and conventional
interactions, and thereby
accommodate the highest energy events.
However, most of the new proposals come with their own peculiar 
difficulties\cite{noexotics,PS96}, so new ideas are still welcomed.

\subsection{A Possible Resolution to the Puzzle}
The purpose of this article is to revisit a neutrino--annihilation 
mechanism which may produce photons and baryons with with 
energies above the GZK cutoff, and to calculate the number of 
photons and nucleons which would arrive at earth with super--GZK 
energies and interact in our atmosphere.
In 1982 it was shown\cite{me} that the annihilation channel of 
cosmic--ray neutrinos on relic neutrinos
should produce an absorption dip in the high--energy cosmic neutrino flux 
if the target neutrinos, the relics predicted by hot Big Bang
cosmology, have a mass large compared to their 
relic temperature $T_{\nu}\sim 10^{-4}$ eV.
Scattering of an ultrahigh energy neutrino
on the nonrelativistic relic neutrino background resonantly 
produces a $Z$--boson at the energy
$E_{\nu_j}^R = M_Z^2/2m_{\nu_j} = 4\,({\rm eV}/m_{\nu_j}) \times 10^{21}$ eV
($j$ labels the three light neutrino types).
The  width of the resonant energy for annihilation is
$\delta E_{\nu_j}^R/E_{\nu_j}^R 
\sim 2\delta M_Z/M_Z \sim 2\Gamma_Z/M_Z = 0.06$.
The annihilation process converts neutrinos with energy within 0.03
of the peak resonant energy $E_{\nu_j}^R$ into highly energetic $Z$--bosons.
The $Z$ produced in each neutrino annihilation immediately decays
(its lifetime is $3\times 10^{-25}\;{\rm s}$ in its rest frame).
70\% of the $Z$ decays are hadronic, with a final state 
known to include on average fifteen neutral pions  and 
1.35 baryon--antibaryon pairs\cite{PDG}.
The fifteen $\pi^0$'s decay to produce thirty high--energy photons.
The $Z$ is highly boosted, with 
$\gamma_Z =E_{\nu}/M_Z \sim 10^{10}$ for super--GZK energies.
Consequently, its decay products are relativistically beamed
into a cone of half--angle $\theta \sim 1/\gamma_Z \sim 10^{-10}$.
We will refer to the collimated high--energy 
end product of this $Z$ production and hadronic decay as a ``$Z$--burst.'' 
If the $Z$--burst points in the direction of earth and 
occurs close enough to earth so that the nucleons 
may propagate to earth without too much energy attenuation, or the
photons without too much absorption, then 
one or more of the photons and nucleons in the burst 
may initiate a super--GZK air shower at earth,
as illustrated in Fig. 1.

Let us call the distance over which a stable particle 
can propagate without losing more than
an order of magnitude of its energy the GZK distance.
For a photon it is 10 to 40 Mpc, with the exact number 
depending on the strengths of the diffuse radio and infrared background.  
For a proton it is 
$D_{{\rm GZK}}\sim -[\ln(1/10)/\ln(0.8)]\times 6$ Mpc $\sim$ 50 to 100 Mpc.
We seek the probability for resonant annihilation of a cosmic 
ray $\nu$ and a relic ${\bar \nu}$ (or vice versa) into the $Z$--boson, 
within the GZK distance of earth (the ``GZK zone''). 
The rate of $Z$--burst production within the GZK zone will be 
greatly enhanced if the relic neutrinos cluster in the potential wells
within the GZK zone\cite{me,roulet}, which includes the wells of 
the Local Supercluster, the Local Group of galaxies,
and of course, our own Galaxy. 
Some clustering is expected if the relic neutrinos
are nonrelativistic, \ie with  mass considerably above their temperature,
$T_{\nu}\sim 2K\sim 10^{-4}$ eV.

The mean energy per hadron from a $Z$--burst 
will be about 40 times less than $E_{\nu_j}^R$,
since the mean multiplicity in $Z$ decay is about 40 \cite{PDG}.
Although the energy per particle is diluted, the 
thirty photons from $\pi^0$--decays and the 2.7 baryons 
in the final state serve to
amplify the propagating particle flux above the GZK cutoff 
by a factor of $\sim 30$.  
We will soon argue that the mean energy per baryon is about ten times 
larger than the mean energy per photon.

Two crucial elements are required for this mechanism to produce  
super--GZK air showers. They are:\\
(i) the existence of a neutrino flux at $\gsim 10^{21}$ eV, and\\
(ii) the existence of a neutrino mass in the 0.1 to 10 eV range.\\
A third ingredient greatly enhances the rate for observed \zb s.
It is:\\
(iii) a significant clustering of the relic neutrino density 
within the GZK zone, e.g. in the potential wells of our Supergalactic 
cluster, Galactic cluster, or possibly even our Galactic halo.\\
We discuss the plausibility of each of these ingredients in turn.

\section{Neutrino Flux, Mass, and Relic Density}
The observations of the highest--energy cosmic rays tell us that some 
primaries are produced at $10^{20}$ eV or more.  The actual
energies of particles at their sources may well be higher than $10^{20}$ eV,
since particles may lose energy while emerging from 
the high--density environment of the source, and, 
as noted earlier, 
nucleons lose energy by scattering on the 2.7K photon background.  
It is likely that whatever 
mechanism produces these most energetic particles also produces charged pions
of comparable energy.  Thus, one may expect neutrino production at 
ultrahigh energy, coming from pion decay and subsequent muon decay\cite{hz}.  
The much smaller opacity for neutrinos as compared to protons in dense sources
such as AGNs makes credible the possibility of  
a neutrino flux considerably above the proton flux at 
highest energies.\footnote
{There is also the possibility that the highest energy neutrinos 
originate in quark jets, 
which themselves result from decay of some supermassive relic particles, 
in which case
the neutrino flux greatly exceeds the proton flux\cite{exparts}.
}

The mean relic neutrino density of the universe is predicted by hot Big Bang 
cosmology.  The density of relic neutrinos with mass below an MeV 
(the neutrino decoupling temperature) is given by a relativistic 
Fermi-Dirac distribution characterized by a single temperature parameter.
The distribution is that of relativisitic neutrinos, even though
the neutrinos are nonrelativistic today, because the distribution
is determined at the decoupling epoch. 
As a result of photon reheating from the era of 
$e^+ e^- \rightarrow \gamma\gamma$ 
annihilation, the neutrino temperature $T_{\nu}\sim 1.95$K 
is predicted to be a factor of
$(4/11)^{1/3}$ less than that of the photon temperature, 
$T_{\gamma}=2.73$K.  
The resulting mean neutrino number density is
$\nuave=(3\zeta(3)/4\pi^2)T^3_{\nu} =54\,{\rm cm}^{-3}$ 
for each light flavor $j$ with an equal number density for
each antineutrino flavor.\footnote  
{If neutrinos are Dirac particles and if 
there are some lepton number violating processes
in the early universe, then the numbers of neutrinos and 
antineutrinos need not be equal; a nonzero chemical potential is 
effectively introduced 
into the neutrino Fermi--Dirac distribution.  
We note that any lepton asymmetry necessarily increases the 
number of neutrinos plus antineutrinos, but we
do not develop this possibility further here.  
For unpolarized Majorana
neutrinos, there can be no chemical potential.}
Note that the predicted relic neutrino density is normalized 
via the temperature relation to the relic
photon density which is measured.  
Consequently, the predicted mean density of 
$\nuave=54\,{\rm cm}^{-3}$ must be considered firm.

The mean energy density in neutrinos is obtained by multiplying together 
mass and mean density and summing over light flavors.  
The result, as a fraction of the closure density, is
$\Omega_{\nu}=\sum_j m_{\nu_j}/(92\,h_{100}^2$ eV),
with $h_{100}$ being the present value of the Hubble parameter 
$H_0$ in units of 100 km/s/Mpc. 
The observational constraint on the Hubble parameter gives 
$0.5\leq h_{100}\leq 1$, with lower values in the range favored.
One sees that a neutrino mass, or
a sum of neutrino masses, up to tens of eV is consistent with 
hot Big Bang cosmology.
Perhaps more importantly, one sees that a neutrino mass around 
or above an eV is required if
neutrino hot dark matter is to contribute in any significant 
way to the evolution
of large--scale structures.  Some have argued that the existence 
of the largest 
observed structures necessitates the existence of hot dark 
matter\cite{primack}.
Furthermore, the simplest explanation for the anomalous 
atmospheric--neutrino flavor--ratio\cite{atmanomaly}
is neutrino oscillations driven by a mass--squared difference of 
$\sim 0.01\,{\rm eV}^2$ \cite{atmososc}, which implies a mass of 
at least 0.1 eV.
The recent LSND measurement is claimed\cite{LSND} to also 
indicate a neutrino mass, of order 1~eV.

\subsection{Gravitational Clustering of Neutrinos}
Some nonrelativistic matter, whether charged or neutral,
is expected to have fallen into gravitational potential wells and ``cooled''.
For charged matter the energy loss mechanism 
is dominantly bremsstrahlung during 
two--body scattering.  For neutral (``dissipationless'') matter,
the mechanism (termed ``violent relaxation'') is many--body gravitational 
scattering wherein some particles gain energy at the expense of 
other particles as a consequence of the time--varying gravitational 
potential\cite{LB67}.
Examples of collapse and capture of dissipationless matter are 
seen in models of gravitational infall\cite{infall}.
A consequence of clustering for massive neutrinos is that 
the mean free path for annihilation is much shorter in 
large--scale potential wells, and the efficiency
for neutrino to hadron conversion is accordingly much higher.
Let us determine the scaling law that governs the increase 
in annihilation rate when neutrinos are clustered
rather than distributed uniformly throughout the universe.
Density scales with clustering length $L$ and
the number of cluster sites $N$ within the volume of the
visible universe $V_{U}$ as $V_{U}/N\,L^3$.
The path length across the neutrino cluster scales as $L/V_{U}^{1/3}$. 
Putting the factors together, and estimating $V_{U}\sim D_H^3$ 
with $D_H\equiv c\,H_0^{-1}$,
the annihilation rate for a neutrino cosmic ray traversing the cluster 
therefore scales as $D_H^2/N L^2$.  
It is useful to write $V_{U}=D_H^3=N\,L^3_{ss}$ to define
the mean distance $L_{ss}$ between neighboring cluster sites.
Then, $N$ may be eliminated to yield the final form $L_{ss}^3/L^2 D_H$ 
of the geometrical scaling factor for annihilation within
a cluster of size $L$ and mean separation $L_{ss}$.  
If the neutrino clustering 
scale is less than the GZK distance of 50 to 100 Mpc,
then the $Z$--burst production rate within the GZK zone also scales 
with the same $L_{ss}^3/L^2 D_H$ law.\footnote
{An average over all cosmic ray trajectories
would show no enhanced annihilation rate.  The fraction of rays that
intersect any relic halo varies as $N L^2$ with clustering, 
to compensate the $1/N L^2$ increase in annihilations within the halos.
The fact that an earthbound observer sees only the $1/N L^2$ 
enhancement factor is due to our ``preferred'' 
vantage point within or nearby a neutrino cluster.
}
We will implement this scaling law shortly.

\section{Neutrino Annihilation on the Relic Background}
For neutrinos with the standard weak interaction, 
there is little hindrance to their propagation over cosmic distances 
except for nonzero mass neutrinos with energy 
near the resonant energy on the $Z$--pole\cite{me,roulet,sommers}.
The annihilation cross section $\sigma_{ann}(\nu_j+{\bar\nu_j}\rightarrow Z)$
is first order in the Fermi constant $G_F$, 
whereas neutrino scattering cross sections are order $G_F^2$.
The relatively large cross section of the resonant process 
is somewhat mitigated by the narrowness of the $Z$ pole 
(2.5 GeV is the width) since it is the energy--integrated cross sections
which determine the annihilation rate of a flux distributed in energy.
The resonant contribution should
dominate the energy--integrated order $G_F^2$ scattering cross 
sections in the energy region of interest, 
particularly if the neutrino flux has a falling spectrum above $E_R$.
In any case, non--resonant scattering contributions can only add to the
resonantly produced super--GZK flux we calculate here.

Define the invariant, energy--averaged annihilation cross section by the 
following integral over the $Z$ pole: 
$\ave \equiv \int \frac{ds}{M_Z^2} \sigma_{ann}(s)$,
with $s$ the square of the energy in the center of momentum frame.
The standard model value for this cross section is 
$\ave= 4\pi G_F/\sqrt{2}=4.2\times 10^{-32}{\rm cm}^2$ for each
neutrino type (flavor or mass basis),
independent of any neutrino mixing--angles since the annihilation
mechanism is a neutral current process.
The energy of the neutrino annihilating at the peak of the $Z$--pole is
$E_{\nu_j}^R = M_Z^2/2m_{\nu_j} = 4\,({\rm eV}/m_{\nu_j}) \times 10^{21}$~eV.
For a given resonant energy $E_{\nu_j}^R$, only neutrinos with the $j^{th}$
mass $m_j$ may annihilate. 
The energy--averaged annihilation cross section $\ave$ is the 
effective cross section for all neutrinos within 
$\frac{1}{2} \delta E_R/E_R =\Gamma_Z/M_Z =3\%$ of their peak 
annihilation energy.
(We will sometimes use $E_R$ generically for resonant energy, as we do here, 
with the understanding that there are really three different 
resonant energies, one for each neutrino mass.)
We will refer to neutrinos with resonant flavor $j$
and with energy in the resonant range 0.97 $E_{\nu_j}^R$ 
to 1.03 $E_{\nu_j}^R$ as ``resonant neutrinos.''
 
\subsection{Rate of Annihilation to $Z$--Bursts}
In a universe where the neutrinos are nonrelativistic but
unclustered 
the mean annihilation length for neutrinos 
at their resonant energy  
would be $\lambda=(\ave \nuave)^{-1} = 4.4\times 10^{29}$~cm.\footnote
{This annihilation length is not too much larger than the 
Hubble size of the universe.
It is this rough equality of lengths which
led to the first observation of the possible absorption dip 
in the neutrino spectrum for neutrinos originating from 
cosmologically--distant sources.
The absorption dip is significantly enhanced by the 
much higher relic densities present in the more--compact 
early universe\cite{me}.}
A cosmic ray neutrino arriving at earth from a cosmically distant source
will have traversed approximately a Hubble distance of space, 
$D_H\equiv cH_0^{-1}=0.9\, h^{-1}_{100} \times 10^{28}$ cm.
In an unclustered relic neutrino sea the annihilation probability 
for such a neutrino at its resonant energy is 
$D_H/\lambda_{ann}= 2.0 \, h^{-1}_{100}\, \%$.
For each 50 Mpc of travel through the mean neutrino density, 
the probability for a neutrino with resonant energy to 
annihilate to a $Z$--boson is $3.6\times 10^{-4}$. 
The branching fraction for a 
$Z$ to decay to hadrons is 70\% .
Consequently, one part in 4000 of the resonant neutrino flux will 
be converted into a $Z$--burst containing 
ultrahigh energy photons and nucleons within 
the 50 Mpc GZK--zone of earth in this unclustered universe.
This value of $0.025\%$ for the probability of a resonant neutrino 
creating a $Z$--burst 
within the GZK zone is the \ul{absolute} \ul{minimum} 
in a Big Bang universe.

\subsection{Neutrino Clustering Included}
With ``local'' clustering of the relic neutrinos, 
the probability for annihilation within the GZK zone is 
significantly enhanced.  The probability is given by that 
found for a Hubble distance of travel
in a non--clustered universe, scaled by the geometric factor 
of Sec.\ 2.1.  The resulting annihilation probability is
$(L_{ss}^3/D_H\,L^2)(2.0\,h^{-1}_{100}\,\% )$,
for a cluster of size $L$ and mean cluster--cluster
separation distance $L_{ss}$.
Let us define some fiducial values for cluster
sizes and mean cluster--to--cluster distances.\footnote
{We realize that a cluster or halo size is not well defined,
but it is useful to pretend that they are for purposes of illustrating
the annihilation enhancement expected from neutrino clustering.
We will see shortly that in fact it is the integrated column density 
of relic neutrinos that determines the annihilation rate. 
Our use of constant densities within well--defined cluster radii 
may be easily translated into column densities.
}
We take $D_S=20$ Mpc for the diameter of the Virgo Supercluster,
the only supercluster well within our GZK zone,\footnote
{Distances from earth to the centers of the 
nearest rich clusters are estimated in \cite{narlikarbook} to be
11, 40, 50, and 60 $h_{100}^{-1}$ Mpc for Virgo, Pisces, Perseus,
and Coma, respectively.
}
and 100 Mpc, typical of the distance across a cosmic void, 
for the supercluster mean separation distance $D_{SS}$. 
We take $D_C=5$ Mpc for the typical diameter of 
galactic clusters and 
$D_{CC}=50$ Mpc as the typical distance between neighboring 
galactic clusters.
For our Local Group of $\sim 20$ galaxies, we take 
$D_{Gp}=2$ Mpc, and $D_{Gp-Gp}=20$ Mpc for the 
mean separation distance between groups.
$D_G=50$ kpc is a typical 
diameter of galactic halos (including our own) and 
$D_{GG}=1$ Mpc is a typical distance between neighboring galaxies.
With these fiducial values for sizes and separation distances,
the respective density enhancements $(L_{ss}/L)^3$
are of order 100, $10^3$, $10^3$, and $10^4$,
for the Supercluster, Cluster, Local Group, and Galactic halo,
assuming that neutrino clustering is more or less as efficient 
as baryonic clustering for these scales. 
The annihilation--probability geometric factors are
$(L_{ss}^3/D_H\,L^2) \sim 0.9$, 1.7, 0.7, and 0.14~$h_{100}$,
respectively, for neutrino clustering on the scales of
the Supercluster, Cluster, Local Group, and Galactic halo,
again assuming efficient neutrino clustering.
Including the 70\% 
hadronic branching ratio of the $Z$, one then gets the 
probabilities 1.3\%, 2.4\%, 1.0\%, and 0.2\% (independent of $h_{100}$),
for $Z$--burst production by a neutrino of relevant flavor at
resonant energy traversing the Supercluster, Cluster, Local Group, 
and Galactic halo, respectively.
Clustering on the small scale of the Galactic halo 
gives the smallest probability.
The larger clusters all give a robust probability,
within a factor of two of each other,
and of order of a per cent.
This is our main result, which we repeat for emphasis:
{\sl the probability for cosmic ray
neutrinos at their resonant energy to annihilate 
within the $\sim 50$~Mpc zone of earth is 
likely within an order of magnitude of 1\%, 
with the exact value depending on unknown aspects
of neutrino mixing and relic neutrino clustering.}

We have assumed here that the neutrino cluster is local, 
in that it contains our earth.  
In this case the scaling law and the
enhanced probability applies for the full sky of cosmic rays.
A ``non--local'' cluster will also yield this enhanced 
probability, but only for cosmic rays with arrival directions 
within the solid angle of the distant cluster.
A detailed calculation of the $Z$--burst rate,
including the possibility of anisotropic neutrino clustering,
will be presented in the next two subsections.

The (in)efficiency of neutrino clustering warrants our attention.
One expects the relic neutrinos to be less clustered than the
baryons, especially on scales as small as the Galactic halo,
for several reasons.  
First of all, neutrinos do not dissipate energy as easily 
as electrically charged protons do.
Secondly, neutrinos have a much larger Jeans (``free--streaming'') length 
than do baryons at the crucial time when galaxies start to grow nonlinearly.
And thirdly, Pauli blocking presents a significant barrier to clustering of
light--mass fermions\cite{TG79}.  As a crude estimate of Pauli blocking,
one may use the zero temperature Fermi gas as a model of the gravitationally
bound halo neutrinos.  Requiring that the Fermi momentum of the neutrinos
not exceed the virial velocity $\sigma\sim\sqrt{MG/L}$ within the cluster, 
one gets 
$\xi=n_{\nu_j}/54\,{\rm cm}^{-3}
\lsim 10^3 (m_{\nu_j}/{\rm eV})^3 (\sigma/200{\rm km s}^{-1})^3$.
The virial velocity within our Galaxy is a couple hundred km/sec,
whereas virial velocities for rich galactic clusters are a
thousand km/s or more.  Thus it appears that Pauli blocking allows
significant clustering on the Galactic scale only if $m_\nu\gsim 1$~eV,
but allows clustering on the larger scales for $m_\nu\gsim 0.1$~eV.
The free--streaming argument also favors clustering on the 
larger scales.\footnote
{An anonymous referee mentioned yet another potential obstacle to
the clustering of neutrinos on the Galactic scale.  It is that the
thermal velocity of relic neutrinos is comparable
to the virial or escape velocity if $m_\nu\lsim 0.1$~eV.
This result is obtained by noting that it is the neutrino's momentum which 
has red--shifted since decoupling, leading to
$m_\nu v_\nu^{\rm Th} = p_D/z_D \approx E_D/z_D \sim 3\,T_D/z_D = 3\,T$,
which implies 
$v_\nu^{\rm Th}\sim 3\,T/m_\nu \sim 5\times 10^{-4}/(m_\nu/{\rm eV})$.
}
This is just as well, for we have shown that 
it is the larger scales of clustering that
give the ${\cal O}(1\% )$ probability for annihilation to
a $Z$--burst.

\subsection{A More Careful Rate Calculation}
The rate calculation may be carried out more carefully.
Let $F_{\nu_j}(E_{\nu},x)$ denote the flux of the $j^{{\rm th}}$
neutrino flavor, as would be measured at a distance $x$ from the source, 
with energy within $dE$ of $E_{\nu}$. 
The units of $F_{\nu_j}(E_{\nu},x)$ are neutrinos/energy/area/time/solid angle.
This flux may be quasi--isotropic (``diffuse''), as might arise from a 
sum over cosmically--distant sources such as AGNs; 
or it may be highly directional, perhaps
pointing back to sources within our Supergalactic plane.
The distinction between diffuse and discrete fluxes
affects the arrival direction of the super--GZK events,
but not the rate calculation.
The change in the neutrino flux due to resonant annihilation is
$d F_{\nu_j}(E_{\nu},x)= -F_{\nu_j}(E_{\nu},x) 
\frac{dx}{\lambda_j (E_{\nu},x)}$,
where the mean annihilation length is 
$\lambda_j (E_{\nu},x) = [\sigma_{ann}(E_{\nu})\, n_{\nu_j}(x)]^{-1}$ 
as before.
$d F_{\nu_j}(E_{\nu},x)/dx$ is therefore the production rate
per unit length of $Z$'s with energy within $dE$ of
$E_{\nu}$, per unit area and unit solid angle.
Integrating the absorption equation over the distance $D$ 
from the emission site to earth gives the depletion 
in the neutrino flux at energy $E_{\nu}$:
\beq
\delta F_{\nu_j}(E_{\nu},D)
= F_{\nu_j}(E_{\nu},0) \left[ 1-\exp\{ -\sigma_{ann}(E_{\nu})\, S_j(D)\} \right],
\label{enabs}
\eeq
with $\delta F_{\nu}(E_{\nu},D)\equiv F_{\nu}(E_{\nu},0)-F_{\nu}(E_{\nu},D)$. 
The relic neutrino column density is defined as
\beq
S(D)\equiv \int^D_0 dx \, n_{\nu_j}(x).
\label{colmdensity}
\eeq
Integrating this equation over neutrino energy then gives the total rate
(per unit area and unit solid angle)
of resonant annihilation, \ie $Z$--burst production, over the distance $D$:
\beq
\delta F_{\nu_j}(D)\equiv \int dE_{\nu} \,\delta F_{\nu_j}(E_{\nu},D)
=\int dE_{\nu} \, F_{\nu_j}(E_{\nu},0)
\left[ 1-\exp\{ -\sigma_{ann}(E_{\nu})\,S_j(D)\} \right].
\label{abs}
\eeq
For a narrow resonance the flux may be evaluated at the resonant energy
and removed from the integral.  
%
%
%
The remaining integral may be made an invariant by using 
$E_{\nu}=s E_R /M^2_Z$ to write
\beq
\delta F_{\nu_j}(D)=
E_R F_{\nu_j}(E_R,0)\int\frac{ds}{M_Z^2}
\left[ 1-\exp\{ -\sigma_{ann}(s)\,S_j(D)\} \right].
\label{abs3}
\eeq
Contact is made with our earlier estimate 
if $\sigma_{ann}(s) S(D)$ is small compared to one.  Then it is sufficient 
to keep the first nonzero term in the Taylor series of the bracketed factor. 
This leads to
\beq
\delta F_{\nu_j}(D) \approx E_R \ave S_j(D) F_{\nu_j}(E_R,0).
\label{inv}
\eeq
where the averaged annihilation cross section is as defined before,
$\ave \equiv \int \frac{ds}{M_Z^2} \sigma_{ann}(s)$, which is equal to
$4.2\times 10^{-32}{\rm cm}^2$.
Eqn.\ (\ref{inv}) shows that for $S(D)\ll 1/\sigma_{ann}(s)$,
the rate for \zb s 
depends linearly on the relic neutrino column density, 
which of course scales as $1/N L^2$, satisfying our scaling law.
For $S(D)\ll 1/\sigma_{ann}(s)$, the attenuation of the neutrino flux
over the distance from the source to the GZK zone will be small,
and we may set $D$ in Eqn. (\ref{inv}) equal to the GZK distance
to get the $Z$--burst rate within the GZK zone.

\subsection{Anisotropic Clustering}
If the neutrino cluster is not isotropic with respect to our
preferred position at earth, the $Z$--burst rate will not be
isotropic either.  The anisotropic rate is easily accounted for
by generalizing the distance $D$ to be vector $\hat{D}$.
The column density integral in Eqn. (\ref{colmdensity}) then 
becomes an integral of $n_{\nu_j}(\hat{x})$ along the vector $\hat{D}$.
There is weak evidence that the super--GZK events may correlate with the 
Supergalactic plane\cite{stanev}.  Such a correlation would arise 
naturally in the model presented here if the SG plane provided either
the super--GZK neutrino flux or the potential well in which neutrinos 
are clustered (or both).

\subsection{Possible Flavor--Mixing of Neutrinos}
It is possible, in fact it is probable, that massive
neutrinos exhibit mixing in analogy to the quark sector of the standard
model of particle physics.  In the mixed case, the flavor states are 
unitary mixtures of the mass states.  Letting $\alpha=e,\mu,\tau$ label 
flavor and $j=1,2,3$ label mass, one writes
$|\nu_{\alpha}>=\sum_j U_{\alpha j} |\nu_j>$.
Then each neutrino flavor at the resonant energy of a given mass state 
has a nonzero probability to annihilate, but with an extra probability factor
of $|U_{\alpha j}|^2$.
For example, the $\nu_\mu$'s and $\nu_e$'s from pion and mu decay will
annihilate at the resonant energy of $m_2$ with the probability factors
$|U_{\mu 2}|^2$ and $|U_{e 2}|^2$, respectively, times what we have
calculated above.\footnote
{The phenomenon of neutrino oscillations will occur
when neutrinos are mixed, and it will affect the flavor populations 
of the cosmic--ray neutrinos. 
However, it will not affect the calculation of
annihilation, because the $\nu-{\bar \nu}$ annhilation 
process requires just a single transformation from flavor to mass basis, 
so the phase differences induced between mass states are not observed.
}
We remark that if the lepton mixing mirrored the known mixing among
the quarks, then the 
heaviest of the three neutrino mass states
should be mainly mixed with the third generation $\nu_{\tau}$.   
If this were the case, then 
$|U_{\mu 3}|^2$ and $|U_{e 3}|^2$ would be small, and
annihilation with a significant probability 
at the lowest resonant--energy $E^R_{\nu_3}$
would arise only if there were a substantial $\nu_{\tau}$ flux 
(of unspecified origin).\footnote
{If $U_{\mu 3}$ and $U_{e3}$ are small, then the vacuum oscillation 
probabilities for $\nu_e\rightarrow \nu_\tau$ and
$\nu_\mu\rightarrow \nu_\tau$ are also small.
}
However, recent data from the SuperKamiokande Collaboration validates
earlier evidence that the $\nu_{\tau}$ and $\nu_{\mu}$ may be near--maximally 
mixed (see ``Note added'' at the paper's end), in which case significant 
$\nu_\mu$ annihilation on $\nu_3$ is expected.
%

We have calculated annihilation probabilities assuming no mixing.  
Without mixing, each flavor--type has a unique
mass, and the resonant energy of each neutrino flavor is unique.
If there is mixing, the factors $|U_{\alpha j}|^2$ can be easily
multiplied in. 

\subsection{Dirac Versus Majorana Neutrinos}
In addition to the flavor--mixing issue, another subtley arises if
the massive neutrinos are of Dirac--type as opposed to Majorana--type.
As the universe expanded and the momenta of the relic neutrinos
red--shifted, they evolved to the
unpolarized nonrelativistic state which they occupy today.  
As a result, if the neutrino is a Dirac particle, then 
the sterile right--handed neutrino and the sterile left--handed 
antineutrino fields are populated equally with the two active fields.
Therefore, for Dirac neutrinos the active densities 
available for annihilation with the incident high energy neutrino are half of
the total densities, and the $Z$--burst production probability is half
of what we quote in this article.  
In contrast, for Majorana neutrinos there are no sterile fields and the
total densities are active in annihilation.  
Majorana neutrinos are favored over Dirac neutrinos in
currently popular theoretical models with nonzero neutrino mass\cite{gelmini}.

\section{$Z$--bursts and the p, n, $\gamma$ Flux Above $10^{20}$ eV}
Having argued that cosmic neutrino/relic neutrino annihilation in our galactic
halo is a possible candidate for $Z$--burst events, 
let us study more closely the particle spectrum contained in the burst.
The decay products of the $Z$ are well--known from the 
millions of $Z$'s produced at LEP and at the SLC.\cite{PDG}.
The respective branching fractions for $Z$--decay into hadrons, 
neutrino--antineutrino pairs,
and charged lepton pairs are 70\%, 20\%, and 10\%.
The mean multiplicity $\enn$ in hadronic $Z$ decay is about 40 particles, 
of which, on average,  17 are charged pions, 9 are neutral pions, 4.4 are
kaons, one is the eta meson, 3.3 are light--quark vector mesons
(mainly the $\rho^0$ and $K^*$'s) which decay to 1.4 kaons, 
0.6 $\pi^0$'s, and 3.2 $\pi^{\pm}$'s, on average;
0.6 are light--quark tensor mesons,
0.8 are $D$ mesons, 0.5 are $B$ mesons,
and importantly, 1.35 are baryon--antibaryon pairs 
which become 2.7 nucleons and antinucleons.
The one eta meson decays on average to 0.8 hard photons (two--photon decay),
and through three--body decays to 
1.2 $\pi^0$'s, and 0.56 $\pi^{\pm}$'s. 
We count the 0.8 hard photons of the eta as 0.4 $\pi^0$ for simplicity.
The 1.3 heavy--quark $D$ and $B$ mesons have many modes available for their
decays.  We estimate their average collective final state as 
1.3 kaons and 3 pions.
The total of seven kaons decay to two-- and three--body 
final states including on average another 
3 neutral pions and 5 charged pions.
The total pion and nucleon count for the $Z$--burst is then,
15 $\pi^0$'s, 28 $\pi^{\pm}$'s, and 2.7 nucleons
(we now mean ``nucleons'' to include the antinucleons as well). 

Among the 15  $\pi^0$'s, we have seen that 
9 are produced directly in $Z$--decay,
while the other 6 arise from decays of various hadronic resonances.
A comparison of the data\cite{PDG} for the momentum spectra 
for direct pions and for protons produced in $Z$--decay reveals
that the proton momentum is about three times that of the direct
pion, on average. Taking into account the differing masses as well,
we estimate that the boosted mean energy per proton is larger than
that of a direct pion by a factor $\sim 3.5$.
The energy of the six $\pi^0$'s produced through resonance decays
may be softer yet.
Weighting the direct and secondary pions appropriately 
then (we take the latter to be softer by $\sim 3$), 
we arrive at a factor of about six for the softness of the mean pion
energy compared to the mean nucleon energy.
Since the photon on average carries half of the parent pion energy,
the mean energy of the $\pi^0$--decay photons in a $Z$--burst 
is expected to be less than that of the nucleons by an order of magnitude.
The energetics arguments we have presented here are qualitative.
The next step would be to quantitatively obtain the energy spectrum
of the super--GZK nucleons and photons by boosting the output of
a simulation program such as ISAJET\cite{ISAJET}, which is
fitted to the final state data in $Z$--decay.

The fate of the charged--pions' decay products is somewhat complicated.
Each charged pion goes through a decay chain which results in an \epm\ and
three neutrinos.\footnote
{The boost factor for the charged and neutral pions is typically 
$\gamma_{\pi}=E_R/40/m_{\pi}=7\,({\rm eV}/m_{\nu})\times 10^{11}$, 
which leads to mean decay lengths $c\,\gamma_{\pi}\tau_{\pi}$
of $5\times \,({\rm eV}/m_{\nu})\,10^{14}$ cm 
$\sim 30\,({\rm eV}/m_{\nu})$ AU 
and $20\,({\rm eV}/m_{\nu})$ km, 
respectively.
The boost factor for the muon in the charged pion decay chain is typically
$0.75\,E_R/40/m_{\mu}=7\,({\rm eV}/m_{\nu})\,\times 10^{11}$,
very similar to the pion boost, 
leading to a mean decay length for the muon of 
$5\times 10^{16}$ cm $\sim 0.02$ pc.
These decay lengths are very short compared to the mean pion and muon 
interaction lengths in the photons and hydrogen
of the intergalactic or even interstellar medium,
so the decays are not impeded by interactions or absorption.
}
At the very least, this decay chain provides a mechanism for injecting
copious $e^{\pm}$'s into any gravitational well rich in relic neutrinos.
Each of the four leptons from the decay chain 
carries on average a quarter of the charged pion's energy.  
The produced neutrinos are below the original 
resonant energy of their parent cosmic ray neutrino,
so they are not relevant for our purposes of creating and 
identifying super--GZK candidates
(we do not consider the possibility 
of a second resonant energy nearby the first).  
The \epm 's are, however, another potential source of super--GZK photons,
via the inverse Compton (IC) process.  
The boost factors for the \epm 's are $\gamma_e\sim E_R/160/m_e\sim 
5\,({\rm eV}/m_{\nu})\times 10^{13}$.
When a photon with incident energy $\omega$ and incident 
angle $\theta$ with respect to the \epm\ velocity is 
scattered through an angle $\alpha$, the final photon energy 
as a fraction of the incident \epm\ energy is
$(1-\cos\theta)(1-\cos\alpha +m_e/2\omega\gamma_e)^{-1}$.
Thus, photons with initial energy satisfying $\omega \gg E_{crit}$, 
with $E_{crit}\equiv m_e/\gamma_e\sim 10^{-8} (m_{\nu}/{\rm eV})$~eV,
will scatter ``catastrophically'' off the \epm 's 
to acquire an energy of order 
$E_R/160\sim 2\,({\rm eV}/m_{\nu})\times 10^{19}$~eV.
$E_{crit}$ is in the radio range, $\sim 10$~MHz
or $\sim 10$~m.  Photons in the 2.73K microwave background have energies
greatly exceeding $E_{crit}$, and may acquire super--GZK energies
from IC--scattering, depending on the value of $m_\nu$.
However, the \epm --photon scattering  cross--section for photons with 
energy exceeding $E_{crit}$ is in the Klein--Nishina regime,
down considerably from the Thomson cross-section, by roughly the factor 
$(E_{crit}/2\omega)\,\ln (2\omega/E_{crit})$.
In contrast to the CMB photons, 
any ambient radio photons with energies far below 
$E_{crit}$ will scatter with the full strength of the Thomson cross section 
($\frac{8}{3}\pi\alpha^2/m^2_e = 6.7\times 10^{-25}{\rm cm}^2$);
however, they will not gain a significant fraction of the \epm 's energy. 
Magnetic fields in the Supergalactic cluster, Local Group, or the
Galactic halo may also degrade the energy of the \epm 's 
via synchrotron emission.\footnote
{Integration of the energy loss formula for relativistic \epm 's
due to synchrotron emission and IC--scattering in the Thomson regime 
gives\cite{eloss}
$E_e(x)=E_e(0) (1+\beta\,E_{20}\,x)^{-1}$, 
with the distance x measured in parsecs, $E_{20}\equiv E_e(0)/10^{20}$ eV, 
and $\beta = 80\,(\rho_{\gamma}+0.6 B^2/8\pi)$.
Here, $\rho_{\gamma}$ is the energy density of the photon background
with energy $\ll E_{crit}$ in units of eV/cm$^3$, 
and $B$ is the ambient magnetic field in units of microgauss.
The electron or positron loses 90\% of its energy in a distance 
$x_{90}=(10/\beta E_{20})$ pc.
Across cosmological distances, there is only an upper limit on $B$ of 
$\sim 10^{-3}$; models typically give $B\sim 10^{-6}$ or less.
However, in clusters of galaxies $B\sim 1$ has been measured,
probably generated by jets from radio galaxies\cite{Kronberg94}.
With $B\gsim 1$, $x_{90}$ is less than a parsec for 
$E_e \gsim 5\times 10^{20}$ eV.
In our Galaxy, $B$ is typically 3, increasing to about 30 when 
averaged over the inner AU of our solar system, and to about $10^6$
within 1000 km of the earth's surface.
The fraction of \epm\ energy transferred to an individual photon through
the synchrotron process is typically small, 
$\sim 10^{-20}\,\gamma_e\,B_\perp $, where $B_\perp $ is the magnetic field
perpendicular to the \epm\ trajectory.
}
A numerical simulation including modeling of the radio, IR and starlight 
backgrounds, and the magnetic fields, may 
be needed to determine the outcome of the energy loss of the \epm 's.  
In what follows, we will conservatively ignore the 
possible contribution to the super--GZK photon flux from 
catastrophic IC--scattering of the 28 \epm 's,
and focus on the 30 $\pi^0$--decay photons 
and the 2.7 nucleons in the $Z$--burst.

In the $Z$--burst, the 30 photons from $\pi^0$--decay and the 2.7 nucleons 
are the candidate primary particles for inducing super--GZK
air showers in the earth's atmosphere.
The 2.7 nucleons may be protons or neutrons.  
With its ten--minute rest--frame lifetime 
enhanced by the boost factor $\gamma_N\sim E_R/\enn m_N \sim
({\rm eV}/m_{\nu})\times 10^{11}$, the neutron will typically travel
$c\tau_n\gamma_N\sim ({\rm eV}/m_{\nu})$ Mpc, 
and then transfer virtually all of its energy to a proton when it decays.
The {\it a priori} photon to nucleon ratio is about 10 on average.\footnote
{As just noted above, there is the possibility of even more super--GZK
photons resulting from IC--scattering of the CMB by the $\sim 28$ \epm 's
liberated in the decay of the $\pi^{\pm}$'s.
}
However, the hardness of the nucleon spectrum compared to the
photon spectrum mitigates this ratio if a selection is made
for the very highest energy particles.
Moreover, the differing attenuation reactions and attenuation lengths of
nucleons and photons will affect the obervable event ratio. 
Also, the development of photon--initiated air--showers 
at $E\sim 10^{20}$~eV is skewed by the LPM effect\cite{Klein98}
and by high--altitude photon--absorption on the earth's magnetic 
field\cite{PS96}, which may affect their identification and measurement.
Finally, the average values for the multiplicities and energies 
presented here must be used with some caution,
since fluctuations in multiplicity and  particle-types per event, 
and in energy per individual particle, are large.\footnote
{Simulations on an event--by--event basis are possible with Monte Carlo
programs such as ISAJET; an additional layer of reality may be added
by also simulating the development of the electromagnetic cascade
in the cosmic medium.
}

\section{Further Signatures from $Z$--bursts}
The particle spectrum in $Z$--decay and the Lorentz factor of the $Z$, 
$\gamma_Z = E_R/M_Z = M_Z/2 m_{\nu} =
0.9\,(m_{\nu}/{\rm eV})^{-1}\times 10^{11}$,
determine the possible signatures of \zb s.  
Going beyond simple shower--counting above the GZK cutoff,
we comment on some of the possible $Z$--burst characteristics:\\
(i) The $Z$--decay products which in the $Z$ rest frame lie within the 
forward hemisphere 
are boosted into a highly--collimated lab--frame cone of half--angle 
$1/\gamma_Z= 2 (m_{\nu_j}/{\rm eV})\times 10^{-11}$ radians.
$Z$--bursts originating within $20 ({\rm eV}/m_{\nu_j})$ parsecs of earth,
if directed toward the earth, arrive with a transverse spatial spread of 
less than one earth diameter.
It is possible for the decay products of a single not--to--far distant 
$Z$--burst to initiate multiple air showers.  
A large area surface array (\eg the Auger project\cite{Auger}) 
or an orbiting all--earth observing satellite 
(\eg the OWL proposal\cite{OWL}) 
could search for these nearly coincident showers.\\
(ii) The mean number of baryon--antibaryon pairs per hadronic $Z$--decay 
is 1.35.  Baryon number conservation requires each hadronic $Z$ decay to 
contain an integer number of baryon--antibaryon pairs.  
If the baryon and antibaryon both strike the 
earth's atmosphere, but are sufficiently separated in arrival time, 
repeater events may be observed.
If the number of baryon pairs per hadronic shower is governed
by Poisson statistics, then the probabilities for 0, 1, 2, 3, 4, and 5
pairs are 26\%, 35\%, 24\%, 10\%, 4\%, and 1\%, respectively.
In estimating the difference in arrival time of two nucleons, it may be 
useful to work with the rapidities of the baryon and antibaryon, 
since rapidity is an additive variable under boosts.
The rapidity difference $\Delta y \equiv y-{\bar y}$ is invariant
under boosts.  
The difference in arrival times of two baryons is 
$\Delta t =D_Z \,(\tanh y -\tanh {\bar y})$,
where $D_Z$ is the distance of the $Z$--burst from earth;
the difference in arrival times of a photon and baryon is
$\Delta t =D_Z \,(1 -\tanh y)$.
Correlated baryon--antibaryon rapidity distributions
may be sought in terrestrial $Z$--decay data.  
If QCD implements baryon number conservation locally, then $\Delta y$ will
generally be small.\\
(iii) The energy of the \zb s are fixed at 
$4\,({\rm eV}/m_{\nu_j}) \times 10^{21}$ eV  
by the neutrino mass(es).  
The energy of individual particles produced in the burst 
can approach this value but cannot exceed it.  
This may serve to distinguish the
$Z$--burst hypothesis from some recent speculations for super--GZK 
events based on SUSY or GUT--scale physics\cite{exparts}, 
in which cutoff energies are expected to be much higher.\\
(iv) From the highest super--GZK event energy $E^{{\rm max}}$,
one can deduce an upper bound on the neutrino mass of
$m_{\nu}< M_Z^2/2E^{{\rm max}}=4\;(10^{21}{\rm eV}/E^{{\rm max}})$ eV.
Similarly, from the mean energy $<\! E \!>$ of super--GZK events
one can estimate the mass of the participating neutrino flavor 
via $m_{\nu}=M_Z^2/2E_R \sim M_Z^2/(2 \enn <\! E \!>) 
\sim 0.5\,(10^{20}{\rm eV}/<\! E \!>)$ eV;
if there is a selection bias toward events at higher energy, then this 
formula gives a lower bound on the neutrino mass.\\
(v) There could be a ``neutrino pile--up'' at two to three decades 
of energy below $E_R$.
These pile--up neutrinos are the result of the hadronic decay chain 
which includes
$Z\rightarrow \,\sim 28\,(\pi^{\pm}\rightarrow \nu_{\mu}+\mu^{\pm}\rightarrow
\nu_{\mu}+\bar\nu_{\mu}+\nu_e/\bar\nu_e+e^{\pm})$;
\ie each of the 70\% of the resonant neutrino interactions which yield 
hadrons produces about 85 neutrinos with mean energy
$\sim E_{\pi}/4 \sim E_R/160$.
These neutrinos are in addition to the neutrinos piling up
from the decay of pions photo-- produced by any super--GZK 
nucleons scattering on the 2.7K background.\\
(vi) A very interesting issue is to what extent the boosted 
$Z$--decay products will contain a copious amount of observable 
brehmsstrahlung gamma--rays. 
Even radio and infrared brehmsstrahlung becomes observable after 
boosting with $\gamma_Z\sim 10^{11}$.
For example, a $10^{-5}$ radio photon becomes an MeV gamma--ray, 
and a $10^{-2}$ eV infrared photon becomes a hard GeV gamma--ray. 
The glib statement that ``the 1/E brehmsstrahlung singularity produces
photons with such low energy that they cannot be observed'' may 
hold for terrestrial accelerators, but does not hold for $Z$--bursts.
A careful field theory analysis of the extreme infrared in $Z$--decay 
is required to address this issue\cite{Zhadronization}. 
It seems possible that \zb s may generate short duration 
gamma--ray bursts observable at earth.  
The hadronization time when boosted to the lab frame, 
$\gamma_Z\times {\rm fm}/c \sim 10^{-12}$ s, 
sets the basic time--scale for particle emission
and brehmsstrahlung.

There are a few more remarks that should be made concerning the 
hypothesis under discussion:\\
(i) A common problem that arises when trying to explain the super--GZK 
air showers with new physics concerns the smoothness of the 
event rate as a function of energy.
Both the event rate and the event energy above the GZK cutoff follow 
extrapolations from below the cutoff.\footnote
{There is statistically weak evidence for a possible event gap at 
energies just above the GZK cutoff\cite{gap}.  
There is also weak statistical evidence for the pile--up of
cosmic rays just below the GZK cutoff.  
If either phenomenon is confirmed in the future, then the case for
new physics is strengthened.}
Smooth extrapolations are not necessarily expected when 
new physics is invoked to explain the super--GZK events.
The proposal in this article makes use of standard model particles and 
interactions only, and so is not automatically subject to this criticism.
If the product of the super--GZK neutrino flux per flavor, 
times the annihilation probability within the GZK zone 
(which we have estimated to be $10^{-2\pm 1}$), 
times the photon and nucleon multiplicity per burst ($\sim 30$),
is comparable to the highest--energy proton flux,
then a smooth extrapolation in event rates results.
Such an occurrence does not seem unnatural.\\
(ii) To be more quantitative, the observed rate of super--GZK events
can be used to predict the cosmic neutrino flux at the resonant energy,
within the context of the neutrino--annihilation hypothesis.
Let ${\cal L}(>E_{{\rm GZK}})$ be the luminosity of super--GZK events
(in units of events/area/time/solid angle).  The relation between this event luminosity 
and the neutrino flux is
${\cal L}(>E_{{\rm GZK}})=E_R\,F_{\nu_j}(E_R)\times 10^{-2\pm 1}\times (\sim 30)$,
which implies
$F_{\nu_j}(E_R)=10^{0.5\pm 1}\times E_R^{-1}{\cal L}(>E_{{\rm GZK}})$.
Because the high energy cross--section is several orders of magnitude below the
hadronic cross--section at these energies, this neutrino flux is not directly
measureable in any neutrino detectors presently under development.\\
(iii) The highest--energy neutrino cosmic--ray flux should point back to 
its sources of origin, and the super--GZK event arrival directions 
should point back to these same sources.
However, if the highest--energy neutrino flux is diffuse, then the 
super--GZK event directions should correlate with the spatial 
distribution of the relic neutrino density.  
The solid angles subtended by any nearby halos may offer
preferred directions for super--GZK events.
As discussed in the earlier section on neutrino clustering,
the Supergalactic plane may be the most probable cluster domain.
Perhaps the angular distribution of
super--GZK events can be used to perform neutrino--cluster tomography.
(We remark that significant flattening of dark halos is suggested
by observation\cite{obsflat} and by numerical simulations\cite{simflat}, 
and that a flattened halo implies 
a larger number density by a factor of the axes ratio.)\\
(iv) If the super--GZK events are due to neutrino annihilation on relics 
as hypothesized here,
and if the high--energy neutrino flux is eventually measured,
then an estimate of the relic neutrino column density out to 
$D_{{\rm GZK}}\sim 50$ Mpc may be made: the column density of 
the annihilating neutrino flavor out to $D_{{\rm GZK}}$ is 
$S^{\nu_j}_{{\rm GZK}}\sim {\cal L}(>E_{{\rm GZK}})/
[\ave F_{\nu_j}(E_R)\,\delta E_R]
=4.5\times 10^{32} {\cal L}(>E_{{\rm GZK}})/ 
[E_R F_{\nu_j}(E_R)]\,{{\rm cm}}^{-2}$.
If $F_{\nu_j}(E)$ is measured below the resonant energy,
an estimate of the neutrino column density can still be made by
extrapolating the flux to $E_R$.
For example, if a power law is assumed with a spectral index $\alpha$,
then  $F_{\nu_j}(E_R)=(E/E_R)^{\alpha} \,F_{\nu_j}(E)$.\\
(v) If the cosmic sources of the highest--energy neutrino fluxes 
are themselves located in potential wells containing bound relic neutrinos,
then the neutrinos produced at the resonant energy may well annihilate 
on the relic neutrinos in the source halo.  
In such a case, a depleted flux of resonant energy neutrinos 
would emerge from the source. 
However, the recessional velocity due to the Hubble flow will red--shift 
neutrinos with energies just above the 
resonant value at the source to the resonant energy in our vicinity.
Any source with cosmological red--shift $z_s$ above 
$z=\delta E/E_R=2\Gamma_Z/M_Z \sim 0.06$ (0.03)
will provide our Galaxy with a full (half) 
flux of neutrinos at the resonant energy.
If the spectrum of neutrinos from a single source can be measured,
one might see the depletion at $E_R$ due to absorption in our halo,
and a depletion at $(1-z_s)\, E_R$ due to absorption in the source halo.

\section{Conclusions}
In summary, if one or more neutrino mass is
of order 0.1 to 1 eV, and if there is a sufficient flux of cosmic ray
neutrinos at $\gsim 10^{21}$ eV, then 
$\nu_{{\rm cr}}+\bar\nu_{{\rm relic}}$
(or vice versa) $\rightarrow Z\rightarrow hadrons\rightarrow nucleons$
and $photons$ in our Supergalactic cluster, Local Group, or Galactic 
halo may be the origin of air shower events above the GZK cutoff.
An abundant number of possible signatures to validate or invalidate this hypothesis 
were presented in this work.
The weakest link in the hypothesis is probably the large neutrino flux
required at the resonant energy, well above the GZK cutoff.
Such a flux severely challenges conventional source models in two ways:
first of all, source dynamics must be able to produce the neutrino flux 
above $10^{20}$~eV,
and secondly, any concomitant photon flux must not violate existing upper limits.
The super--GZK photon flux produced in non--local $Z$--bursts is also a potential 
liability, for it will cascade down to produce an abundance of gamma--rays
which must not violate existing upper bounds.

A logical next step toward validating or invalidating the hypothesis is a
numerical simulation of the details of the model\cite{propgnsim}.
The more pressing details amenable to simulation 
are the energy spectrum of the super--GZK 
nucleons and photons, the cluster densities of the relic neutrinos,
and the flux of gamma--rays resulting from casacading of the highest--energy
photons produced in the $Z$--bursts and at the neutrino source.
If future simulations and the inevitable increase in event statistics validates 
the hypothesis, then highest--energy air--showers are our window to
the relic neutrino gas liberated from the primordial 
early--universe plasma when the universe was only one second old!\\
\vspace{0.25cm}\\
{\bf Acknowledgements:}
This work is supported in part by the U.S. Department of Energy grant 
no.\ DE-FG05-85ER40226.
Some of this work was performed at the Aspen Center for Physics.
This paper has benefitted from discussions with Francis Halzen
and Venya Berezinskii.
I thank Stuart Wick for a careful reading of the manuscript.\\
\vspace{0.25cm}\\
{\bf Notes added:}\\
1. As this manuscript was being completed, a related work appeared on the net,
astro-ph/9710029, by D. Fargion, B. Mele and A. Salis.\\
2. Two recent experiments have updated the 
neutrino ``anomalies'' which are most
simply interpreted in terms of oscillations of neutrinos having nonzero mass.
The neutrino mass values which emerge from these data analyses are within the 
range of interest for producing $Z$-bursts at energies just above the GZK cutoff.
The SuperKamiokande experiment finds that the
neutrino--oscillation explanation of the anomalous flavor--ratio of atmospheric neutrinos
requires\cite{SKam:atm}
$\nu_\mu\to\nu_\tau$ oscillations 
(or $\nu_\mu\to\nu_s$ oscillations to a sterile neutrino $\nu_s$) 
with a mass--squared difference of
$0.5\times 10^{-3} \lsim \delta m^2_{\rm atm} \lsim 6\times 10^{-3}\rm\ eV^2$
and mixing angle $\sin^2 2\theta_{\rm atm} \gsim 0.8$.
This implies a roughly 50\% overlap in probability for the $\nu_\mu$ with
a mass eigenstate having a mass of {\it at least} 0.02~eV. 
The LSND experiment has additional evidence for $\nu_\mu$--$\nu_e$ flavor 
conversion, which continues to support the inference of
a mass--squared difference in the
range $0.3$~eV$^2 < \delta m^2_{LSND} < 10$~eV$^2$, and a small mixing probability
(although the KARMEN experiment\cite{KARMEN} rules out part of the LSND allowed region).
If this LSND result is validated with future experiments, it implies
some overlap of both $\nu_\mu$ and $\nu_e$ with a mass eigenstate
having a mass of {\it at least} 0.5~eV. 
%
%
%

%
\vfill\eject
\noindent
{\bf Figure Caption:}\\
{\bf Fig. 1.} Schematic diagram showing the production of a $Z$--burst
resulting from the resonant annihilation of a cosmic ray neutrino on a 
relic neutrino.  If the $Z$--burst occurs within the GZK zone ($\sim 50$~Mpc)
and is directed towards the earth, then photons and nucleons
with energy above the GZK cutoff may arrive at earth and initiate
air showers.
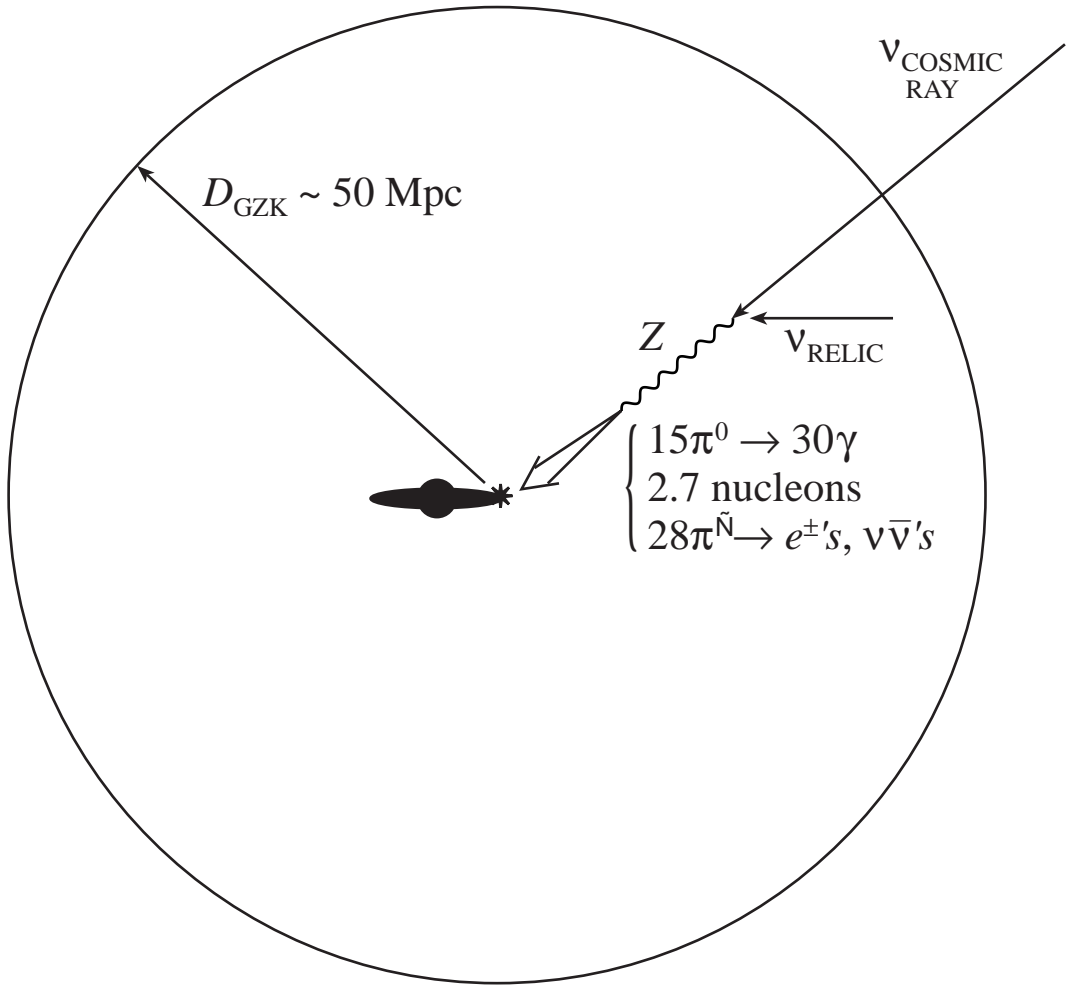
\end{document}